\documentstyle[emulateapj]{article}

\def\etal{{\rm et~al.\ }}
\def\hmpc{\;h^{-1}{\rm Mpc}}

\def\hkpc{h^{-1}{\rm kpc}}

\def\junits{{\rm \;keV\;cm^{-2}s^{-1}sr^{-1}keV^{-1}}}

\newcommand{\PSbox}[3]{\mbox{\rule{0in}{#3}\includegraphics{#1}\hspace{#2}}}

\lefthead{Dav\'e \etal} 
\righthead{The Warm-Hot Intergalactic Medium}

\begin{document}

\title{ Baryons in the Warm-Hot Intergalactic Medium }

\author{
Romeel Dav\'{e}$^{1,2}$, Renyue Cen$^{1}$, 
Jeremiah P. Ostriker$^{1}$,
Greg L. Bryan$^{3,4}$, 
Lars Hernquist$^{5}$, Neal Katz$^{6}$, 
David H. Weinberg$^{7}$,
Michael L. Norman$^{8}$,
and Brian O'Shea$^{8}$
} 
 
\footnotetext[1]{Princeton University Observatory, Princeton, NJ 08544;
 rad,cen,jpo@astro.princeton.edu}
\footnotetext[2]{Spitzer Fellow}
\footnotetext[3]{Department of Physics, Massachusetts Institute of Technology, 
Cambridge, MA 02139; gbryan@alcrux.mit.edu}
\footnotetext[4]{Hubble Fellow}
\footnotetext[5]{Harvard-Smithsonian Center for Astrophysics, 
Cambridge, MA 02138; lars@cfa.harvard.edu}
\footnotetext[6]{Department of Astronomy, 
University of Massachusetts, Amherst, MA, 01003; nsk@kaka.phast.umass.edu}
\footnotetext[7]{Department of Astronomy, 
Ohio State University, Columbus, OH, 43210, dhw@astronomy.ohio-state.edu}
\footnotetext[8]{Astronomy Department, University of Illinois at 
Urbana-Champaign, Urbana, IL 61801; norman,bwoshea@ncsa.uiuc.edu}

\begin{abstract} 

Approximately $30-40\%$ of all baryons in the present day universe
reside in a warm-hot intergalactic medium (WHIM), with temperatures
between $10^5<T<10^7$~K.  This is a generic prediction from six
hydrodynamic simulations of currently favored structure formation
models having a wide variety of numerical methods, input physics,
volumes, and spatial resolutions.  Most of these warm-hot baryons
reside in diffuse large-scale structures with a median overdensity
around $10-30$, not in virialized objects such as galaxy groups or galactic
halos.  The evolution of the WHIM is primarily driven by shock heating
from gravitational perturbations breaking on mildly nonlinear,
non-equilibrium structures such as filaments.  Supernova feedback
energy and radiative cooling play lesser roles in its evolution.  WHIM
gas is consistent with observations of the 0.25~keV X-ray background
without being significantly heated by non-gravitational processes
because the emitting gas is very diffuse.  Our results confirm and
extend previous work by Cen \& Ostriker and Dav\'e et al.
\end{abstract}
 
 
\section{Introduction}

Observations indicate that most of the baryonic matter in the universe does
not reside in galaxies.  At high redshifts ($z\ga 2$), the overwhelming
majority of baryons are in a diffuse, photoionized intergalactic medium
(IGM), observable as \ion{H}{1} absorption lines in the spectra of
distant quasars (\cite{cen94}; \cite{zha95}; \cite{mir96}; \cite{her96}).
The baryonic density inferred from these observations (\cite{rau97})
is in good agreement with nucleosynthesis arguments based on observed
deuterium abundances (\cite{tyt96}).  But by redshift zero the total
baryonic component, inferred from \ion{H}{1} absorption, gas and stars in
galaxies, and other observations, has declined to a level small compared
to that seen at high redshift and expectations from nucleosynthesis
(\cite{fuk98}; \cite{hog99}).  Thus the question arises, where are the
baryons at the present epoch?

By the current epoch, hierarchical structure formation has produced
deep potential wells into which the baryons accrete, thereby moving a
significant portion of the baryons from the IGM into stars, galaxies,
groups, and clusters.  These complex evolutionary processes may now be
modeled directly using cosmological hydrodynamic simulations, enabling
an investigation into the location and phase of baryonic constituents
in the present-day universe, and suggesting possible avenues for their
direct detection.  

These hydrodynamic simulations of structure formation indicate that
baryons in the universe reside in four broad phases, defined by their
overdensity $\delta\equiv\rho/\bar\rho-1$ (where $\bar\rho$ is the mean
baryonic density) and temperature $T$:
\begin{enumerate}
\item {\it Diffuse:} $\delta<1000$, $T<10^5$~K.  Photoionized intergalactic
gas that gives rise to Lyman alpha absorption.
\item {\it Condensed:} $\delta>1000$, $T<10^5$~K.  Stars and cool galactic gas.
\item {\it Hot:} $T>10^7$~K.  Gas in galaxy clusters and large groups.
\item {\it Warm-Hot:} $10^5<T<10^7$~K.  The ``Warm-Hot Intergalactic Medium" (WHIM), discussed here.
\end{enumerate}.

Cen \& Ostriker (1999; hereafter \cite{cen99}) and Dav\'e \etal (1999;
hereafter \cite{dav99}) predicted that a sizeable fraction of all
baryons at the present epoch reside in this last warm-hot phase (see
Figure~2 of \cite{cen99} and Figure~12 of \cite{dav99}).  Such a reservoir
has significant implications for producing an accurate census of baryons
for comparison with nucleosynthesis arguments (e.g. \cite{fuk98}) because
gas at these temperatures and densities is difficult to detect in either
absorption or emission, as we will discuss in \S\ref{sec: disc}.


\begin{deluxetable}{ll|cccccc|cccc|cc}
\footnotesize
\tablecaption{Simulation parameters.}
\tablewidth{0pt}
\tablehead{
\colhead{} &
\colhead{Code} &
\colhead{$\Omega$} &
\colhead{$\Omega_\Lambda$} &
\colhead{$n$} &
\colhead{$\Omega_b$} &
\colhead{$H_0$} &
\colhead{$\sigma_8$} &
\colhead{$L$ \tablenotemark{a}} &
\colhead{$\epsilon$ \tablenotemark{b}} &
\colhead{$m_{\rm bar}$ \tablenotemark{c}} &
\colhead{Physics \tablenotemark{d}} &
\colhead{$\Omega_{\rm WHIM}\over \Omega_b$} &
\colhead{$C_{\rm WHIM}$ \tablenotemark{e}}
}
\startdata
D1 & PTreeSPH & 0.4 & 0.6 & 0.95 & 0.0473 & 65 & 0.8 & 50 & 7 & $8.5\times 10^8$ & 1,3,4 & 0.30 & 244 \nl
D2 & PTreeSPH & 0.4 & 0.6 & 0.95 & 0.0473 & 65 & 0.8 & 11.11 & 3.5 & $1.1\times 10^8$ & 1,3,4 & 0.29 & 405 \nl
C1 & TVD-PM   & 0.37& 0.63& 0.95 & 0.049  & 70 & 0.8 & 100 & 200 & $1.6\times 10^8$ & 1,2,3,4 & 0.42 & 34 \nl
C2 & TVD-PM   & 0.37& 0.63& 0.95 & 0.049  & 70 & 0.8 & 50 & 100 & $2.0\times 10^7$ & 1,2,3,4& 0.37 & 106 \nl
B1 & AMR      & 0.3 & 0.7 & 1.0  & 0.04   & 67 & 0.9 & 100 & 50 & $7.9\times 10^9$ & adiabatic  & 0.32 & 208 \nl
B2 & AMR      & 0.3 & 0.7 & 1.0  & 0.04   & 67 & 0.9 & 100 & 1 & $9.9\times 10^8$ & 1,2,3,4 & $\approx 0.3$\tablenotemark{\dag} & $\approx 400$\tablenotemark{\dag} \nl
\enddata
\tablenotetext{a}{Box size in comoving $\hmpc$.}
\tablenotetext{b}{Spatial resolution in comoving $\hkpc$; 
for PTreeSPH and AMR, this is the highest resolution achieved in dense regions.}
\tablenotetext{c}{Baryonic mass resolution in $M_\odot$.  In TVD-PM, this is 
the average mass per cell.}
\tablenotetext{d}{$1\equiv$~H, He cooling; $2\equiv$~Metal cooling; $3\equiv$~Photoionization; $4\equiv$~Star formation \& feedback.}
\tablenotetext{e}{Clumping factor of warm-hot gas at $z=0$; see \S\ref{sec: emission}.}
\tablenotetext{\dag}{Values extrapolated from $z=0.75$ to $z=0$ based on
a comparison with simulation B1.}
\end{deluxetable}

In this paper we study the nature and evolution of warm-hot gas in a
$\Lambda$CDM universe, using a suite of cosmological hydrodynamic
simulations having a wide range of numerical and physical parameters.
The purpose of this paper is to ask how robust simulation predictions
are to these parameters, to examine the physical state of warm-hot gas
in the universe, and to investigate constraints on warm-hot gas from
soft X-ray background observations.

In \S\ref{sec: sims} we briefly describe the simulations used.  The
evolution of diffuse gas at high redshift into condensed, hot, and WHIM
gas at the present epoch is quantified in \S\ref{sec: evol}.  Our
primary result, presented in \S\ref{sec: whim}, is that the WHIM
accounts for a significant fraction ($\sim 30-40\%$) of baryonic mass
at $z=0$, regardless of variations in spatial resolution, input
physics, or hydrodynamic algorithm.  This is because the evolution of
the WHIM is driven primarily by shock-heating of gas falling into
gravitationally-generated potential wells, a process which is
well-understood and modeled, and only secondarily by processes such as
supernova feedback, radiative cooling, and photoionization.
Furthermore, most of the WHIM gas is at relatively low overdensities,
so shock heating of intergalactic gas occurs during flows onto
non-equilibrium large-scale structures such as filaments.  The majority
of warm-hot gas is found outside of virialized objects such as galactic
halos and galaxy groups.  The low overdensities explain why radiative
cooling and supernova heating do not drive its evolution, and why the
presence of this component probably does not violate constraints from
the X-ray background, as we show in \S\ref{sec:  emission}.  Finally,
we briefly discuss strategies for direct detection of this gas, noting
that the easiest place to detect emission from WHIM gas is relatively
close to galaxies, where it is dense, even though most of the warm-hot
baryons are not in these regions.  In summary, the WHIM is a robust and
generic prediction of currently popular hierarchical structure
formation models, and it contains roughly one-third of all baryons in
the universe today.

\section{Simulations}\label{sec: sims}

We use six cosmological hydrodynamic simulations of randomly-selected
volumes in $\Lambda$-dominated Cold Dark Matter universes, employing
three different numerical techniques, with a range of physical and
numerical parameters.  These parameters are summarized in Table~1.
Simulations run with Parallel TreeSPH (\cite{dav97}) are labeled D1 and
D2, simulations run with TVD-PM (\cite{ryu93}) are labeled C1 and C2,
and simulations using Adaptive Mesh Refinement (AMR; \cite{bry99})
are labeled B1 and B2.  The cosmology chosen is close to the
``concordance model" which is in agreement with a wide variety of
observations (\cite{bah99}).

For our purposes, the significant inputs are the variations in spatial
resolution ($1\rightarrow 200\hkpc$), baryonic mass resolution
($2\times 10^7\rightarrow 10^9 M_\odot$), boxsize ($11.11\rightarrow
100\hmpc$), input physics, and hydrodynamic algorithms.  D1 and D2 are
high spatial resolution Lagrangian (particle-based) simulations, C1 and
C2 are lower spatial resolution Eulerian simulations having high mass
resolution and employing the Total Variation Diminishing scheme, and B1
and B2 are high resolution adaptive mesh simulations based on the
Piecewise Parabolic Method (note the exceptional resolution of $\sim
1\hkpc$ achieved by AMR in the high density regions).  All simulations
except B1 include radiative cooling from H and He, photoionization
heating, and star formation; B1 includes none of these.  C1, C2, and B2
additionally include metal-line cooling, with the metallicity
determined self-consistently from supernova output.  All simulations
use a $\Lambda$-dominated cold dark matter universe, having similar
power at cluster and galaxy scales.  B2 has only been evolved to
$z=0.75$, but this will be sufficient to indicate the relevant trends.

\section{Evolution of Intergalactic Gas}\label{sec: evol}

Figure~1 shows the evolution of the baryonic mass fraction in the four
phases described above.  The four panels show results from simulations
D1, D2, C1 and C2.  Despite differences in simulation volume,
resolution, and numerical method, the evolution of various phases is
qualitatively similar.  At high redshift ($z\ga 2$), the dominant
fraction of baryons resides in diffuse gas (dashed lines) giving rise
to Lyman alpha forest absorbers, as has been explored in detail
elsewhere (see \cite{rau98} for a review).  As structure forms, diffuse
gas is shock-heated, producing warm-hot gas (solid lines).  Gas that is
driven to higher densities due to gravitational instability is able to
cool into the condensed phase (dotted lines) and form stars.  At lower
redshifts ($z\la 2$), large potential wells are produced that
shock-heat gas to $T>10^7$~K, giving rise to hot cluster gas
(dot-dashed curves).  Nevertheless, at the present epoch, the total
fraction of baryons in clusters is small.

\PSbox{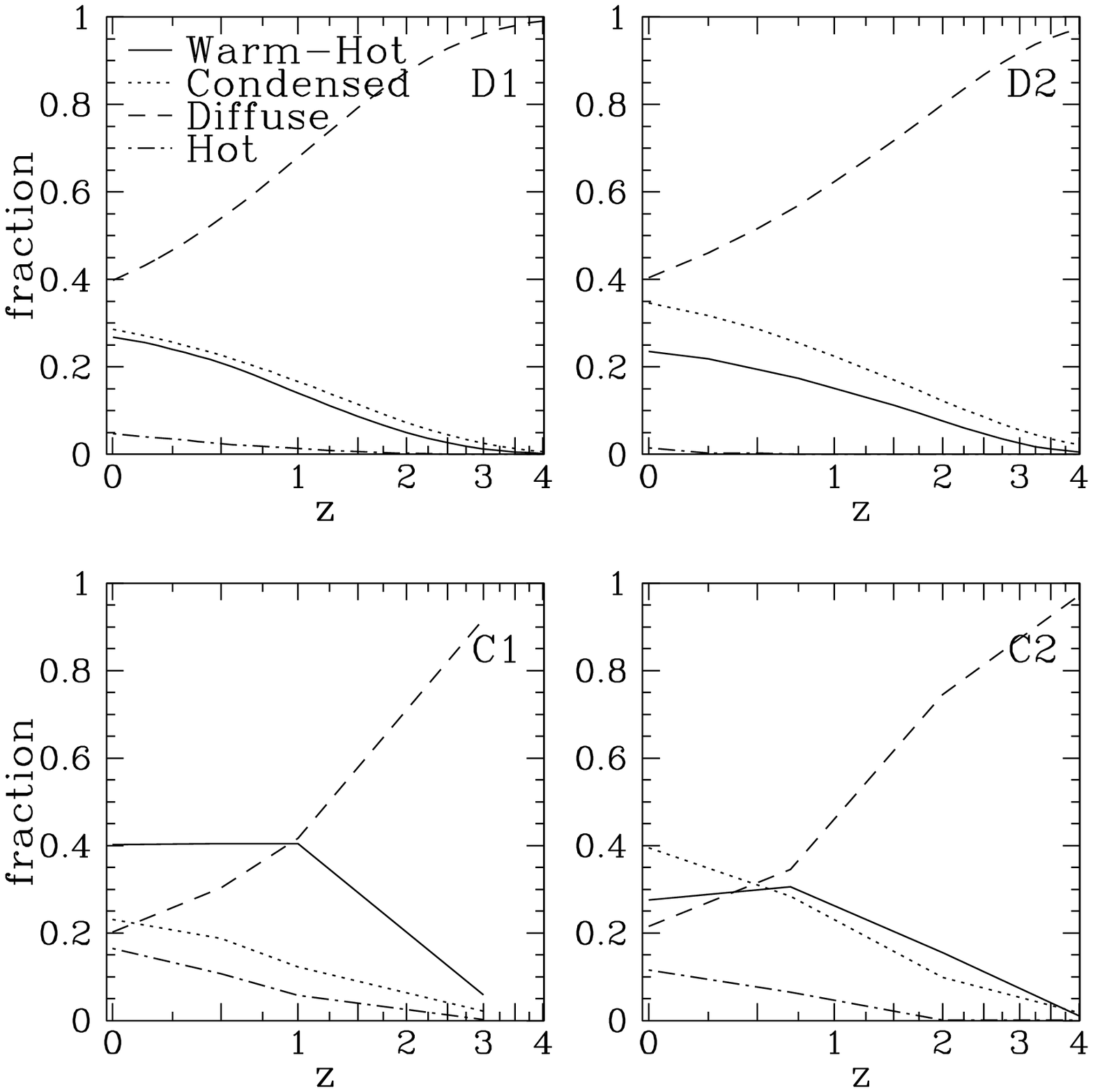 angle=0 voffset=-110 hoffset=-40 
vscale=50 hscale=50}{3.0in}{3.5in} 
{\\\small Figure~1: Evolution of mass fractions in four baryonic phases,
in four simulations.
\vskip0.1in
}

Figure~1 shows that the fraction of diffuse baryons at the present
epoch is between 20\% and 40\%, the fraction of warm-hot baryons is
$30-40\%$.  Gravitationally bound baryons, i.e. those in stars,
galactic gas, clusters, and intragroup media, make up the rest.  Thus
baryons in the present-day universe are divided roughly equally among
diffuse, warm-hot, and bound components.

While all the simulations are qualitatively similar, the exact
distribution of gas in these phases is sensitive to details of the
simulations.  For instance, the condensed phase fraction is sensitive
to how gas cools and forms stars in these simulations, which in turn is
significantly affected by resolution (since cooling and star formation
~$\propto\rho^2$).  Also, the effect of supernova feedback heating
depends strongly on resolution, as we will discuss later.  The growth
of structure is affected by the amount of large-scale power present.
In particular, the hot gas fraction in clusters is sensitive to cosmic
variance, since it is dominated by the largest virialized objects in the
volume.  Other phases, including the warm-hot phase, are less
sensitive to volume effects, as we show in \S\ref{sec: owhim}.

All things considered, it is not surprising that there are differences
up to a factor of two in the fractions in various phases at $z=0$.
While the differences may be significant, an investigation of their
exact causes is beyond the scope of this work, though these differences
offer clues into the physical processes driving WHIM evolution, as we
will explore in \S\ref{sec: owhim}.  Rather, we focus on the remarkable
qualitative consistency in the evolution of various gas phases, given
the variety of simulation methodologies utilized.

For the condensed phase, a census of baryons in stars and cold gas
estimates its mass fraction to be around 20\% (\cite{fuk98}).  The
simulations shown generally produce somewhat higher values for the
condensed fraction, but there are uncertainties in the contributions
from low-mass stars and supernova remnants.  Of these simulations, C2
contains the largest fraction of condensed gas ($\sim 40\%$), while the
rest are all lower, down to $\sim 25\%$ for simulation C1.  B1, of
course, has virtually no condensed gas since it does not include
cooling.  In what follows, we will fix the condensed fraction to be
20\% and redistribute the excess condensed gas equally among all the
other components, assigning a somewhat arbitrary fraction of 1/3 of it
to the WHIM (the only component we examine from here on), in order to
facilitate a less resolution-dependent comparison of the intergalactic
components in these simulations.

\section{The Warm-Hot Intergalactic Medium}\label{sec: whim}

\subsection{Evolution of $\Omega_{\rm WHIM}$}\label{sec: owhim}

\PSbox{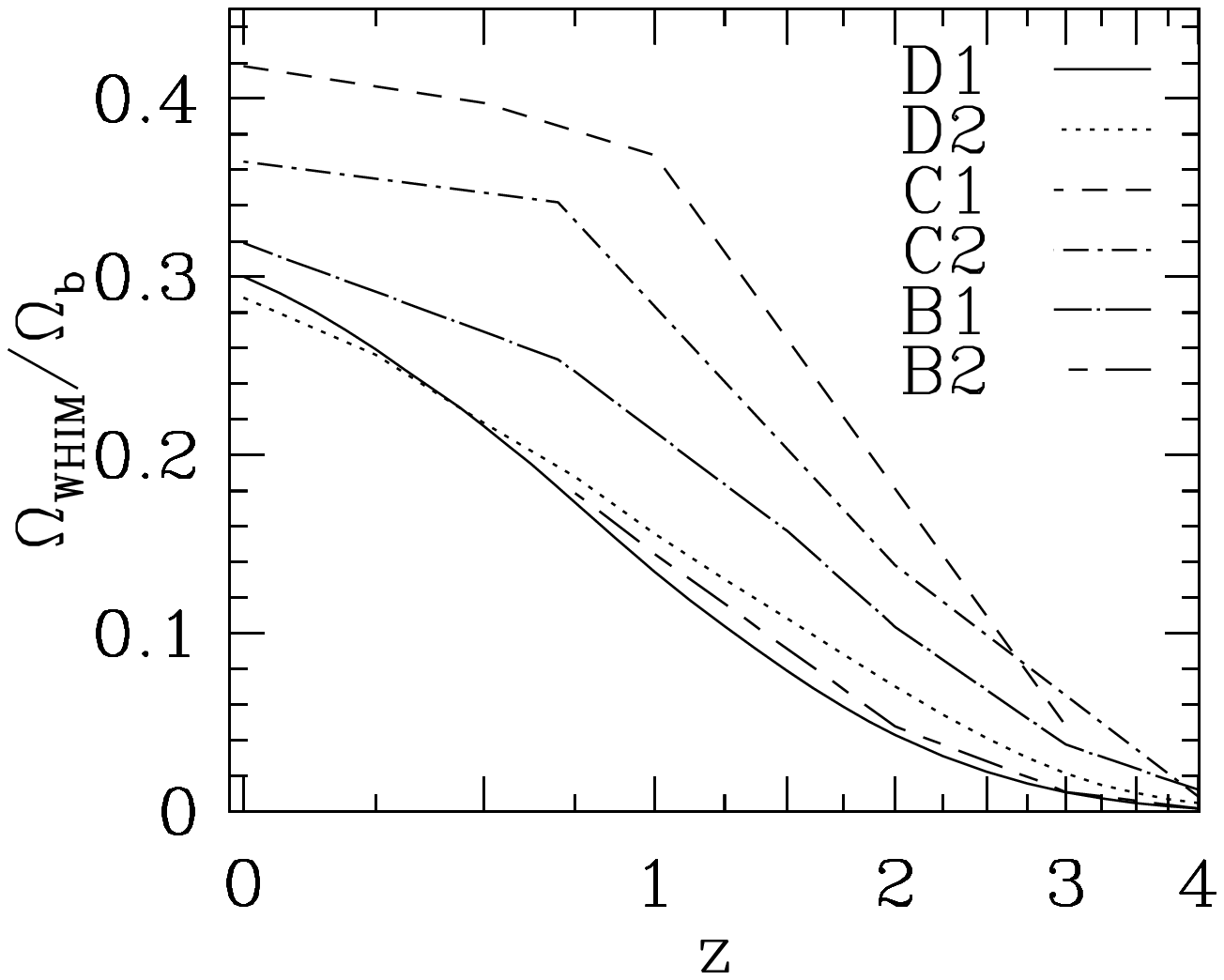 angle=0 voffset=-210 hoffset=-40 
vscale=50 hscale=50}{3.0in}{2.2in}
{\\\small Figure~2: Evolution of $\Omega_{\rm WHIM}$ 
in our six simulations.  For this comparison, the baryonic fraction
in cold galactic gas and stars has been fixed at 20\% in all simulations.
\vskip0.1in
}

Figure~2 shows the evolution of the mass fraction of baryons in warm-hot gas,
$\Omega_{\rm WHIM}/\Omega_b$, for our six simulations, with the condensed
component having been fixed at 20\%.  C1 and C2 have the highest WHIM
fractions, at 42\% and 35\% of baryonic mass, respectively, at $z=0$.
D1 and D2 have lower WHIM fractions at the present epoch, around $\sim
30\%$.  B2 has only been evolved to $z=0.75$, but its evolution closely mirrors
that of the Lagrangian runs, so we expect its WHIM fraction will also
be $\sim 30\%$ by $z=0$.  These values are listed in Table~1.  By
the present epoch, all simulations have WHIM fractions within 50\%
of each other.  Simulation B1 does not include radiative cooling, so it
is an unphysical model that will only be used for comparison with B2.

We expect that the evolution of warm-hot gas is governed by shock
heating of intergalactic gas onto large-scale structure, supernova
feedback, and radiative cooling.  Numerical considerations such as
resolution, volume and hydrodynamic algorithm may also play a role.  In
this section we use case-by-case comparisons among our six simulations
to examine how each of these processes affects WHIM evolution in our
models.

All of our simulations explicitly include the growth of structures and the
accretion of gas onto those structures.  The qualitative consistency of
WHIM evolution in all our simulations, including simulation B1 without
cooling or star formation, suggests that gravitational shock heating of
gas falling on large-scale structures is the dominant heating mechanism
for WHIM gas.  This is an important point, as it suggests that all
other differences between these simulations are of secondary
importance.

The most obvious differences between models are that D1, D2 and B2 all
predict similar WHIM evolution, while C1 and C2 have more WHIM gas.
The primary distinction between these sets of simulations is spatial
resolution.  This manifests itself in various ways, as it affects gas
cooling, the rate at which stars form, and the the injection of
supernova heat energy into intergalactic gas.  To elaborate on this
last point, high-resolution runs (D1, D2, B2) deposit feedback energy
(thermal only) locally in very high density regions where stars are
forming, and thus it quickly radiates away, with virtually none of it
being distributed into the intergalactic medium.  Conversely,
lower-resolution runs (C1 and C2) deposit the same supernova energy
over hundreds of kiloparsecs, resulting in a significant fraction of
feedback energy escaping into diffuse regions where it cannot
radiatively cool away.  None of the simulations here are capable of
resolving supernova-driven galactic winds through a multi-phase
interstellar medium, which are likely to be responsible for distributing
energy and metals into the diffuse IGM (\cite{mac00}; \cite{efs00}),
thus we are relying on heuristic modeling of these processes.  As
we discuss in \S\ref{sec:  phys}, it is in the distribution of
feedback energy where resolution plays its most crucial role in predicting
the evolution of WHIM gas.

First we examine the effect of radiative cooling.  Simulation B1
has no cooling, whereas simulation B2 is a similar run with cooling
(and is our highest spatial resolution run).
Cooling has a greater effect at earlier times because intergalactic gas
is denser then and thus can cool significantly.  After $z\sim 3$, the
rate of growth of the WHIM fraction is similar in B1 and B2, suggesting
that WHIM gas is no longer affected by cooling.  Even by $z=0.75$, the
difference between B1 and B2 is not large, indicating that cooling plays
a minor role in the overall evolution of WHIM gas.

Eulerian codes resolve shock fronts better than SPH in diffuse regions
(\cite{kan94}), because they have a higher density of resolution
elements there, and, in the case of the Eulerian codes discussed here,
because they incorporate explicit shock capturing algorithms to resolve
fronts over two cells.  If a significant component of shock heating
arises from small-scale shocks unresolved by SPH, Eulerian codes may
produce higher temperatures.  One might suspect that the difference
between PTreeSPH runs and TVD-PM runs could be partially due to this
effect.  However, B2 is also an Eulerian simulation, yet it has nearly
the same WHIM fraction as the Lagrangian runs.  So this cannot be a
significant effect.

Simulation volume could also play a role in the amount of WHIM gas,
since larger volumes contain larger perturbations that can result in
stronger gravitational shock heating.  This may explain the difference
between C1 and C2, whose box lengths differ by a factor of two.
However, D1 has a box length five times that of D2, yet their WHIM
fractions are nearly identical at all times.  Furthermore, B2 has twice
the box length of D1, yet its WHIM fraction is in good agreement with
D1 and D2.  As we will show in the next section, typical WHIM gas is at
very moderate overdensities, and cosmic variance in that regime is
typically small, in contrast to the hot IGM fraction which is dominated
by rare, massive objects (i.e. clusters) whose numbers are more
sensitive to simulation volume.

It is possible that opposing effects are causing D1, D2 and B2 to be
similar.  For instance, D1 resolves shock fronts better than D2,
perhaps resulting in more shock heating and making up for its lack of
large-scale power.  It is also possible that the adaptive refinement of
B2 results in more cooling in shock fronts, compensating for its larger
volume.  However, unless all of these effects are of secondary
importance, cancellations at the level we find here would require a
remarkable coincidence.

The parameters of the underlying cosmological model are expected to
have a non-negligible effect on the WHIM component.  The rate of
structure evolution in large part determines how much shock-heated gas
is present at any epoch.  While we have not sampled a range of
cosmologies here, \cite{dav99} examined four cosmologies, namely
$\Lambda$CDM, Tilted CDM, Cold+Hot DM, and Open CDM models, using
numerical parameters similar to those of simulation D2.  The differences
between the WHIM fractions (called ``shocked" gas in \cite{dav99}) in
those models at the current epoch is comparable to the differences seen
here due to other factors, as indicated by their Figure~12.  Similarly,
\cite{cen99} examined $\Lambda$CDM, Cold+Hot DM, and Open CDM models,
and also found broad consistency.  Thus while our quoted fractions may
be specifically for a $\Lambda$CDM cosmology, our qualitative
conclusions are unlikely to be highly sensitive to cosmology.

In summary, the evolution of $\Omega_{\rm WHIM}$ is qualitatively
consistent among all simulations examined, and results in $\approx
30-40\%$ of baryons residing in the WHIM today.  Radiative cooling,
simulation volume, and algorithmic details do not significantly affect
WHIM evolution in these runs.

\PSbox{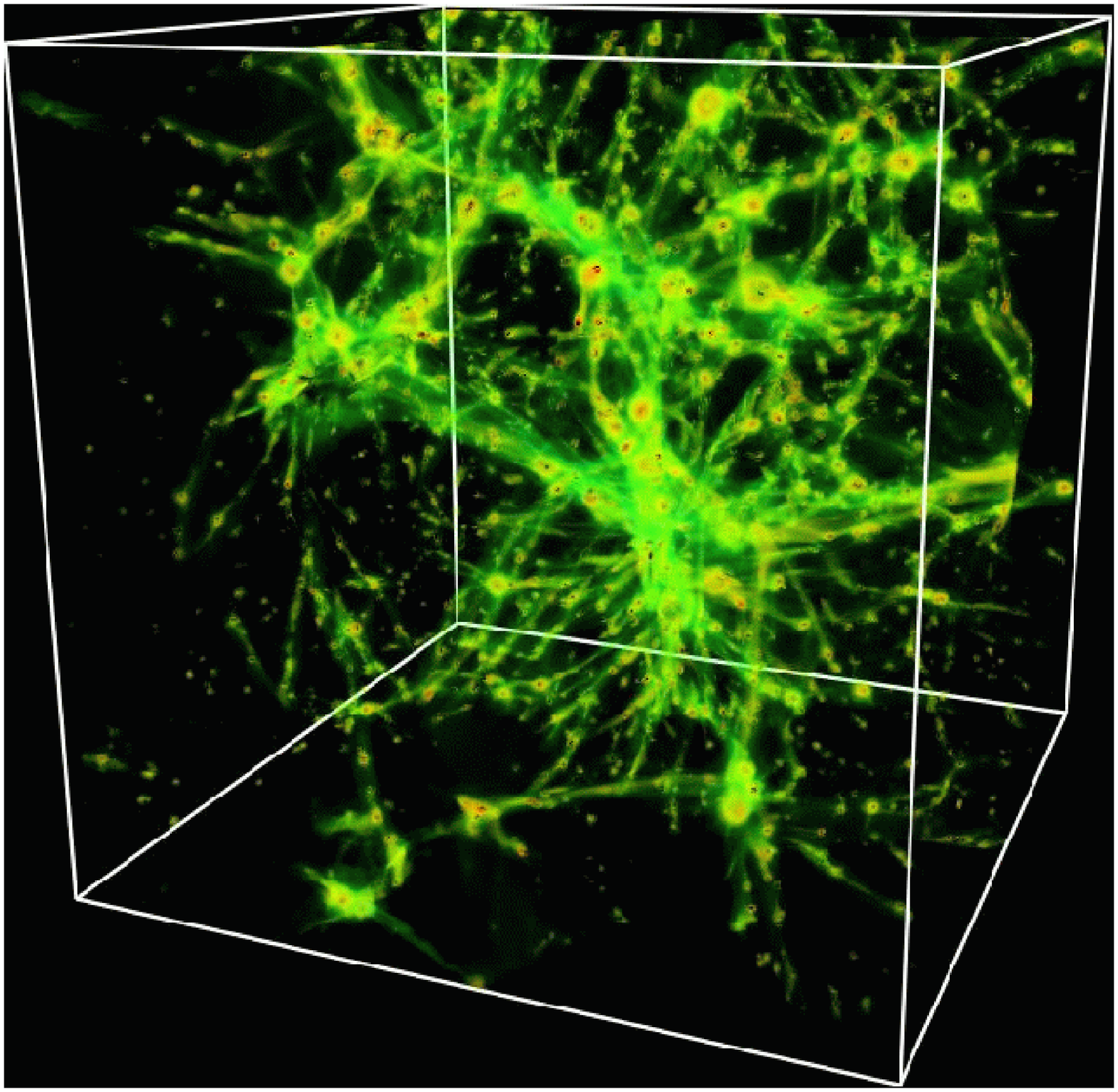 angle=0 voffset=-20 hoffset=0 
vscale=38 hscale=38}{3.0in}{3.6in} 
{\\\small Figure~3:  WHIM gas in simulation C2.  Contours are color
coded by overdensity; green represents overdensity $\delta\sim 10$, 
while red shows $\delta \sim 10^4$.
\vskip0.1in
}

\subsection{The Physics of WHIM Gas}\label{sec: phys}

We now explore the physics that drives the formation and evolution of
WHIM gas.  A qualitative physical picture of WHIM gas may be obtained
by examining Figure~3, which shows the location of gas in simulation
C2 having $10^5<T<10^7$K, with contours color-coded by density.  WHIM
gas is seen to primarily trace out filamentary large-scale structures.
Like the intergalactic medium, WHIM gas does cluster around dense
regions that are sites of galaxy formation.  However, we will show
below that the majority of WHIM gas is contained in the filaments.

\PSbox{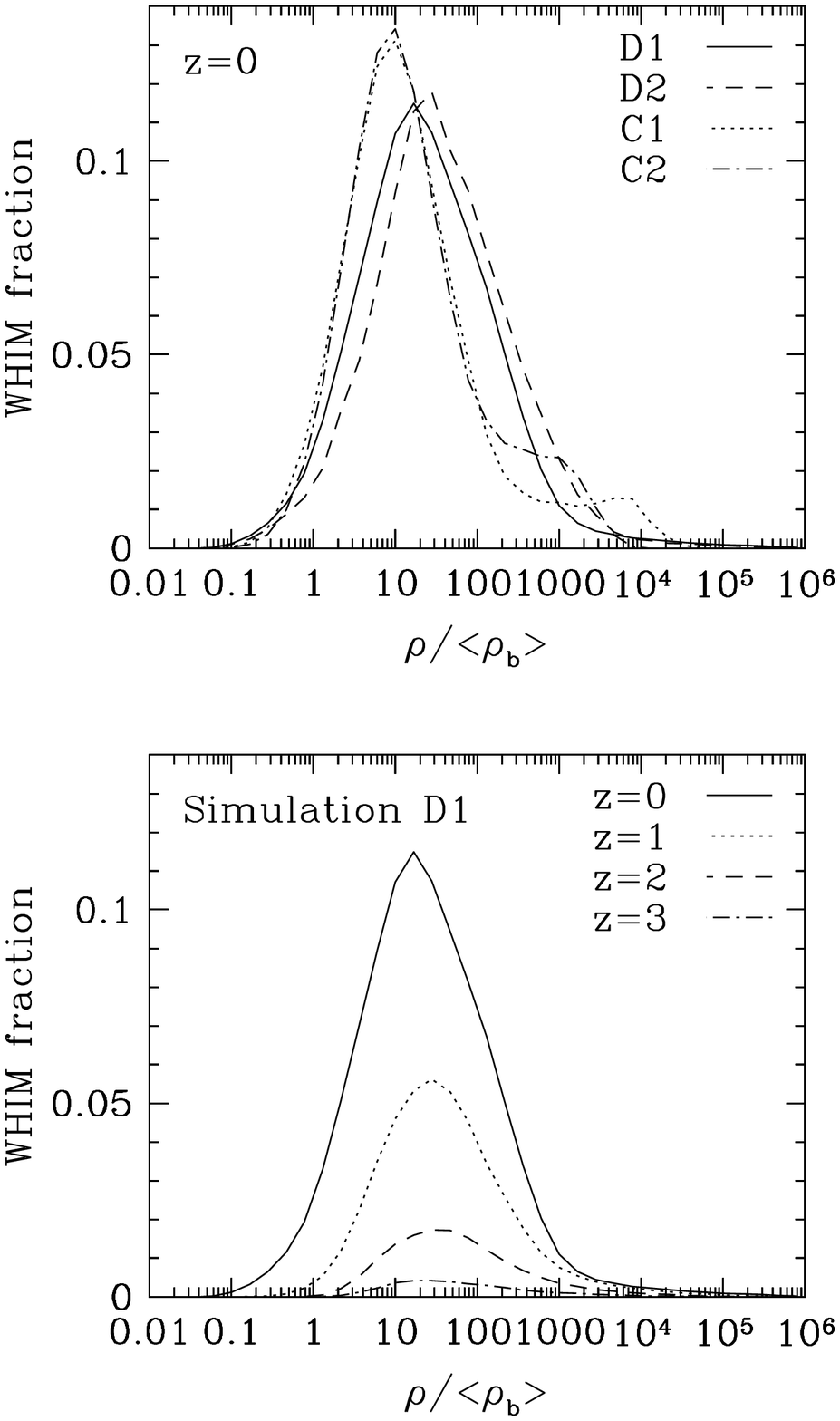 angle=0 voffset=-20 hoffset=0 
vscale=40 hscale=40}{3.0in}{3.8in} 
{\\\small Figure~4:  Top panel shows the mass fraction of WHIM gas as a
function of density at $z=0$ in our simulations.  Bottom panel
shows the same quantity at $z=0,1,2,3$ in simulation D1.
\vskip0.1in
}

The top panel of Figure~4 quantifies the spatial distribution of WHIM
gas in the universe.  It shows a histogram of WHIM gas mass as a
function of density, for simulations D1, D2, C1, and C2.  All four
simulations consistently show that the dominant fraction of WHIM gas is
at relatively low densities, with a peak around an overdensity of $\sim
10-30$.  Simulations B1 and B2 are not shown because B1 does not
include cooling and B2 has only been evolved to $z=0.75$.  Still, B1 at
$z=0$ and B2 at $z=0.75$ show peak overdensities of 18 and 21,
respectively, so they are consistent with the other simulations.
$70-80\%$ of WHIM baryons lie in the overdensity range $5<\delta <200$,
typical of filaments.

The typical overdensities of WHIM gas are much smaller than that of
matter contained in bound, virialized objects.  A maximal estimate of
the bound fraction of WHIM gas may be obtained as the fraction of WHIM
gas with $\delta \ga 60$, which is approximately the overdensity at the
virial radius of an isothermal sphere in a $\Lambda$CDM cosmology.  The
mass fraction of WHIM gas with $\delta \ga 60$ is $\approx 30\%$ in our
simulations.  Clearly some of this gas will not be bound, but rather
infalling material.  As an independent check, we use the group finding
algorithm SKID\footnote{\tt
http://www-hpcc.astro.washington.edu/TSEGA/tools/skid.html} (Spline Kernel
Interpolative DENMAX) to identify bound warm-hot particles in
simulation D1, and find a bound fraction of WHIM gas between $\sim 10\%$
and $\sim 25\%$, depending on the linking length used ($50-500\hkpc$).
The lower end of this range probably corresponds to gas contained in
galactic halos, although the extent of a galactic halo becomes
ill-defined when it resides within a group or cluster.  In summary,
WHIM gas is mostly an intergalactic component, with a majority of it
residing outside of virialized structures such as galaxies or groups.
Note that coronal gas in galaxies also lies in the warm-hot temperature
range, but it is a very small fraction of the galactic baryonic mass, as
most galactic baryons are tied up in stars and cold gas.

How can intergalactic gas be heated to $T>10^5$~K without being bound
in a massive virialized halo?  One way would be if energy was added
from non-thermal processes such as supernova feedback.  However, this
is not the driving process for WHIM gas in these simulations.
Supernova feedback energy is added in all simulations shown in
Figure~4, but in simulations D1, D2 and B2 the effect is negligible,
since, as discussed before, feedback is added purely thermally into
very dense regions, where it radiates away almost immediately and adds
no heat to diffuse regions.  But even in these high-resolution
simulations, where supernovae add negligible heat to WHIM gas, the
typical overdensities are smaller than those typical of virialized
halos.

Supernova heating is, however, likely to be responsible for the somewhat
lower overdensities in simulations C1 and C2 as compared to the higher
resolution simulations, so its effect is non-negligible.  The
differences cannot be due to other spatial resolution effects such as
cooling, as C1 and C2 are themselves quite similar, as are D1 and D2.
Instead, the lower spatial resolution in C1 and C2 results in
significant supernova energy being deposited in intergalactic gas, as
described in the previous section.  This additional feedback heating 
raises the pressure of intergalactic gas and lowers the typical
density.  Furthermore, because supernova energy is not radiated away
immediately, there is a small component of warm-hot gas in and around
galaxies seen as the high-density bump for C1 and C2 in Figure~4.  Note
that the densities in D1 and D2 are computed slightly differently than
in C1 and C2; in the former, Lagrangian runs, the density field is
smoothed by the (variable) smoothing length of the particle, while in
the Eulerian runs the smoothing is fixed at the cell size.  However,
since in both cases the smoothing is done on scales smaller than the
density variations, this is not expected to produce any systematic
differences in the resulting densities.

The effect of supernova heating can be roughly estimated by the
following simple argument.  Consider two extreme cases, one where all
supernova energy is deposited within dense regions where it immediately
radiates away, and another where supernova energy is distributed
uniformly over all baryons in the universe.  The former case is a
reasonable approximation for the Lagrangian runs D1 and D2 as well as
the high-resolution AMR run B2, while the latter is closer to the
Eulerian runs C1 and C2 (though clearly a much more extreme case).  In
the former case, supernovae add no heat to WHIM gas.  In the latter,
the specific heat added per baryon yields a temperature increase of
\begin{equation}
\delta T_{\rm SN} = {\Omega_\ast \; \rho_{\rm crit}\; \epsilon_{\rm SN} \over k_B \; \bar{n}_H },
\end{equation}
where $\Omega_\ast$ is the cosmic mass fraction in stars, $\rho_{\rm
crit}$ is the critical density, $\epsilon_{\rm SN}$ is the specific
energy output per unit mass of stars formed, and $\bar{n}_H$ is the
mean number density of H atoms.  We take $\Omega_\ast=
0.1\Omega_b=0.002 h^{-2}$, and $\epsilon_{\rm SN}=2.5\times
10^{48}$~erg ${\rm g}^{-1} M_\odot$ from a Salpeter IMF with each
supernova from a star with $M>8M_\odot$ outputting $10^{51}$~ergs.
Then, $\delta T_{\rm SN}=0.2$~keV $\approx 2\times 10^6$~K.  Assuming
isentropic heat distribution, $\rho T^\frac{3}{2}=$~constant, so gas
with $T\approx 4\times 10^6$~K (the maximum of the temperature
distribution, as we will show in Figure~5) will have its density
reduced by roughly a factor of two compared to the no heating case.
This is roughly the level of reduction seen in the Eulerian as compared
to the Lagrangian runs.  Clearly this model is overly simplistic, as it
would predict that all intergalactic gas has $T\ga 10^6$~K, but it does
roughly indicate the magnitude of the effect of supernova heating.

Note that there are no other heating processes in these simulations
that contribute significantly to the WHIM.  For instance,
photoionization only heats gas to $\sim 10^4$~K.  Any other heating
process one could devise, such as cosmic rays or supernovae occuring in
dwarf galaxies in voids, would add more pressure support to the gas,
and thus would push the typical density of WHIM gas even lower.  Thus our
simulations indicate that WHIM gas is heated to $T>10^5$~K primarily by
shock-heating of gas accreting onto large-scale structure.  These
structures, typically filamentary, are {\it not} virialized or in
dynamical equilibrium.

In the previous section we showed empirically that radiative cooling
has a minor effect on WHIM evolution.  This may be understood
physically given that warm-hot gas is typically at such low
overdensities that it cannot collapse into virialized objects, and due
to the metagalactic photoionizing background most intergalactic gas
lies well within the optically thin regime (see \cite{dav99},
Figure~10).  Thus typical WHIM gas is too diffuse to self-gravitate or
self-shield, and has no way to achieve the densities required to make
radiative processes important.  Improving the numerical resolution of
our simulations would not change this result, as it is based on simple
physical arguments.

The bottom panel of Figure~4 shows a histogram of WHIM gas mass for the
D1 model at $z=0,1,2,3$ (solid, dotted, dashed, long dashed lines).
There is a slight trend to higher overdensities at earlier times, since
at early times the contribution from virialized structures is greater.
This is because gas has not had time to accrete and shock on more
diffuse structures, and the largest structures at earlier times have
lower temperatures that can fall into the warm-hot range.  Still, this
is a minor effect; basically, the peak overdensity does not evolve
significantly with redshift.

\PSbox{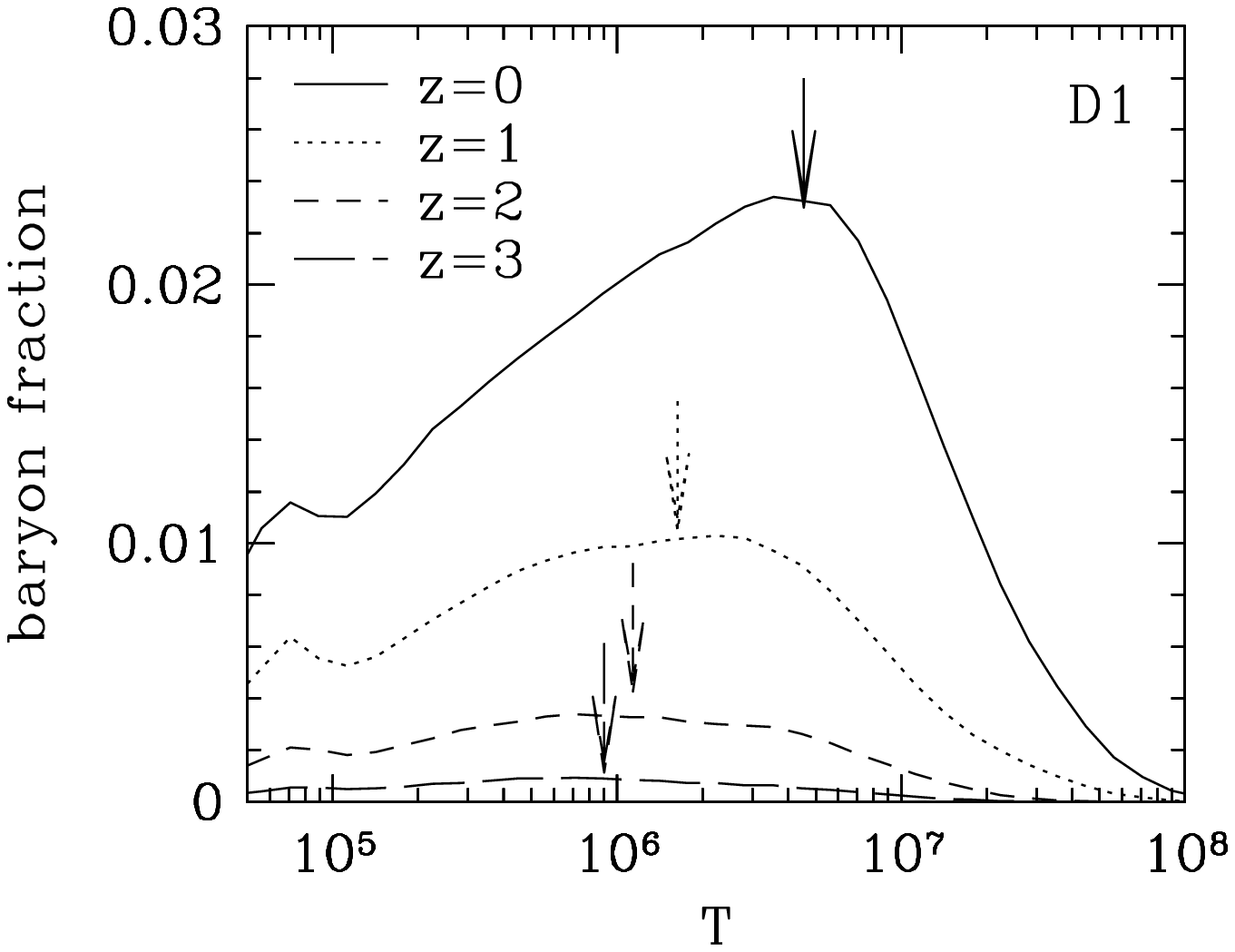 angle=0 voffset=-160 hoffset=-10 
vscale=40 hscale=40}{3.0in}{1.9in}
{\\\small Figure~5:  The mass fraction of baryons as a
function of temperature in simulation D1, at $z=0$ (solid), $z=1$ (dotted),
$z=2$ (short dashed) and $z=3$ (long dashed).  The arrows
indicate the predicted peak temperature from gravitational shock heating
at various $z$, from equation~\ref{eqn: tpeak}.
\vskip0.1in
}

Figure~5 shows the temperature distribution of intergalactic gas with
$T>10^{4.5}$~K in simulation D1 at $z=0,1,2,3$.  The amount of gas at
these warm-hot and hot temperatures grows with time, as seen from
Figures~1 and 2.  The temperature at the peak of the distribution grows
in time as well, reflecting the fact that the universe contains hotter
structures at later times.  The peak temperature at all redshifts falls
within the warm-hot range, and by $z=0$ it is up to $\sim 4\times
10^6$~K.  Note that this is close to the temperature of the excess
diffuse emission seen by \cite{wan93} in ROSAT data.  Figure~5 also
shows that the mass of the WHIM component is insensitive to our
somewhat arbitrary choice of $10^5$ K as the defining lower
temperature.  A choice of $10^{4.5}$ K (as in \cite{dav99}) or even
$10^{5.5}$ K would not drastically affect our conclusions.

An analytic estimate of the evolution of the peak temperature can be
obtained by considering the temperature of intergalactic gas shock
heated on mildly nonlinear large-scale structure.  If the length scale
going nonlinear at a given epoch is $L_{\rm nl}$, and the perturbation
collapses on a timescale $t$, the resulting sound speed behind the
shock will be $\sim L_{\rm nl}/t$.  Given that the perturbation has
taken a Hubble time to collapse, $t\sim H^{-1}$, where $H$ is the
Hubble constant at time $t$.  The resulting post-shock temperature is
then (\cite{cen99})
\begin{equation}\label{eqn: tpeak}
T_{\rm nl} \propto c_{\rm nl}^2 = K (H L_{\rm nl})^2\;,
\end{equation}
where $K$ is a constant, at any given epoch.  The value of $T_{\rm
nl}$, with $K=0.3$, is shown at each redshift plotted in Figure~5 by
the arrows above each curve.  This value of $K$ produces roughly the
correct peak temperature at $z=0$, and as can be seen from Figure~5, it
is a reasonable fit to the peak temperature at higher redshifts, though
it becomes progressively more difficult to identify a peak.  This value
of $K$ also produces the correct evolution of the globally averaged
temperature in simulations C1 and C2, as shown in \cite{cen99}.  Thus
the evolution of the temperature distribution of WHIM gas is consistent
with the interpretation that it is heated by gravitationally-induced
shocks on mildly nonlinear large-scale structure.  Since there is a
wide range in the properties of the collapsing structures and therefore
infall velocities, there is also a wide range in gas temperatures.

\PSbox{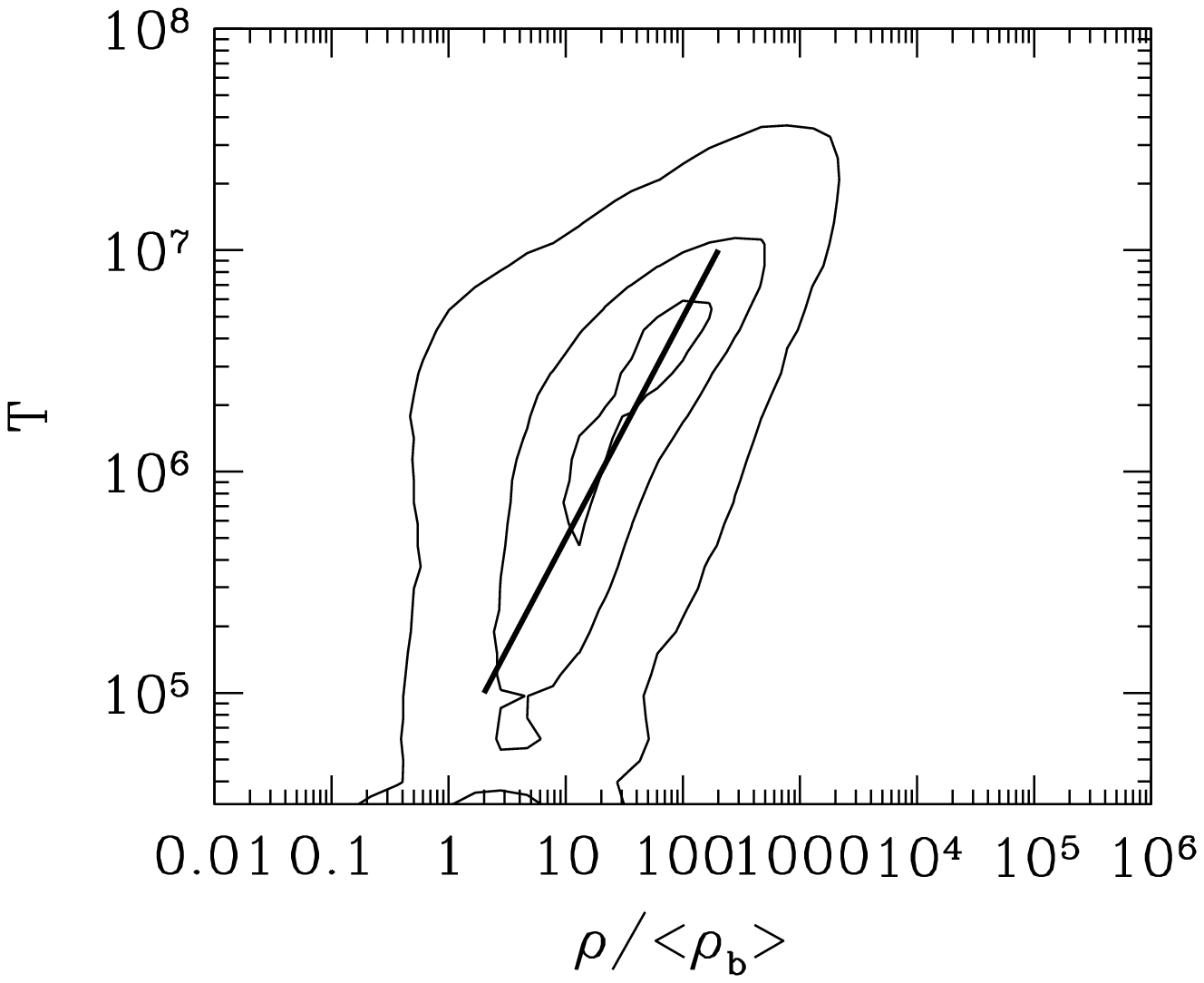 angle=0 voffset=-160 hoffset=-10 
vscale=40 hscale=40}{3.0in}{1.9in}
{\\\small Figure~6:  Contours in temperature and density for
simulation D1 at $z=0$, enclosing 10\%, 50\% and 90\% of the
baryons in the range shown.  Density and temperature are
correlated in the WHIM regime.  Thick line indicates a scaling
of $\rho/\bar\rho_b = T/10^{4.7}$ in the warm-hot temperature range.
\vskip0.1in
}

The temperature and density of WHIM gas are correlated.  Figure~6 shows
a contour plot of mass within the warm-hot range, as a function of
density and temperature, for the D1 model at $z=0$.  The contour levels
enclose 10\%, 50\% and 90\% of the mass in the temperature and density
ranges shown in the plot.  The thick line indicates an ``equation of
state" $\rho\propto T$ that provides a reasonable fit to gas in the
range $10^5<T<10^7$~K.  This relationship is different from that of
diffuse gas, which typically has $\rho \propto T^{1.7}$, and the
temperature-density relation of WHIM gas has much greater scatter.  The
higher temperature, different slope, and greater scatter all reflect
the importance of shock heating as the dominant mechanism controlling
the thermal properties of WHIM gas; the ``equation of state'' for
diffuse gas, on the other hand, arises from the competition between
photoionization and adiabatic cooling due to Hubble expansion
(\cite{hui97}).  Figure~6 also suggests that detecting WHIM gas in
emission will be easier for gas that is at the highest end of the WHIM
temperature range, since that gas will be both denser and hotter.
However, the dominant portion of WHIM gas is at lower temperatures,
which is perhaps most easily detected via absorption lines
(\cite{tri00}).

\section{Constraints From the Soft X-Ray Background}\label{sec: emission}

Gas with temperatures in the range $10^5<T<10^7$~K will emit thermally in
the soft X-ray band.  The extragalactic soft X-ray foreground (SXRB)
flux at 0.1-0.4~keV is roughly $\sim 20-35\junits$ (\cite{war98}),
though uncertainties are large because galactic coronal gas provides an
increasing foreground to lower energies.  At slightly higher energies
($\sim 1$~keV), the XRB has further been resolved nearly completely
($\sim 80-90\%$) into point sources, mostly AGN (\cite{mus00}).
Reasonable arguments then allow only a small contribution to the SXRB
from diffuse gas, $\la 4\junits$ (Wu, Fabian \& Nusser 1999, hereafter
\cite{wu99}).  Such a limit, in principle, places constraints on the
amount of gas in the universe at warm-hot temperatures.

These limits were explored in two independent papers using similar
methodologies, \cite{wu99} and Pen (1999).  Both papers argue that the
standard picture of hierarchical formation of virialized objects
results in a predicted SXRB that exceeds the observed limits.  They
suggest that significant non-gravitational heating, typically $\sim
1$~keV per baryon, is required to unbind warm-hot gas from virialized
objects in order to satisfy the SXRB constraints.  In this section we
discuss these constraints in the context of WHIM gas, and find that our
simulations paint a very different picture for soft X-ray emission than
the simple models assumed in those two papers.  A full calculation of
the SXRB from these simulations is a complicated undertaking (because
of metallicity, bandpass, and numerical issues) that is beyond the
scope of this paper.  However, the physical properties of WHIM gas in
the simulations are quite different from the properties assumed by WFN
and Pen (1998), and we will show that scaling our results to theirs
suggests that the observed SXRB flux does not rule out the WHIM
predicted by these simulations.  The essential difference is one of
density.  Both WFN and Pen (1998) base their calculations on a
\cite{pre74} (1974) analysis, which implicitly assumes that gas is in
virialized objects with typical overdensity $\ga 200$.  However, most
of the WHIM gas in the simulations is in lower-density filamentary
structures rather than virialized objects, thus the SXRB emission is
lower (cf. Figure~5).

We can quantify this difference in typical overdensity by considering
the clumping factor of the emitting gas.  If we define the clumping
factor for gas component $g$ as
\begin{equation}\label{eqn: clump}
C_g\equiv <\rho_g^2>/<\rho_g>^2, 
\end{equation}
then the free-free emissivity from that component is
\begin{equation}\label{eqn: sxrb}
\epsilon_{\rm SXRB} \propto \;\;<\rho_g^2 T_g^{0.5}>\;\; \propto C_g \Omega_g^2 T_g^{0.5},
\end{equation}
where $\rho_g$, $\Omega_g$, $T_g\approx 10^6$~K, and $C_g$ are the
density, mass fraction, temperature, and clumping factor of the gas
emitting in soft X-rays.  The flux $j_{\rm SXRB}$ of soft X-ray
background is then the emissitivity multiplied by path length $\sim
\frac{1}{3} cH^{-1}$.  \cite{wu99} argue, sensibly, that it is
predominantly warm-hot gas ($10^5<T<10^7$~K) that is responsible for
soft X-ray emission.  This means the appropriate clumping factor $C_g$
is that of {\it warm-hot} gas, $C_{\rm WHIM}$.

There are several ways to calculate $C_{\rm WHIM}$ in our simulations.
One can directly calculate it from equation~\ref{eqn: clump}, which is
the approach we use for our Eulerian simulations.  For Lagrangian
simulations, because each particle represents a different volume of
gas, it becomes more numerically convenient to calculate $C_{\rm
WHIM}\approx \xi_{\rm WHIM}(0)$, where $\xi_{\rm WHIM}(r)$ is the
two-point correlation function of WHIM gas at radius $r$.  

The resulting $C_{\rm WHIM}$ values for our simulations at $z=0$ are
listed in Table~1.  All simulations show clumping factors in the range
$\sim 30-400$.  The smaller clumping factors in C1 and C2 arise directly
because the WHIM gas is typically less dense in these models as compared
to D1, D2, B1 and B2 due to greater supernova feedback in the diffuse IGM,
as explained before.

Our clumping factors are significantly less than that used by
\cite{pen99}, who adopts $C\ga 900$ and argues $C\ga 10^4$.  The
reason is that the clumping factor of \cite{pen99} is actually that of
{\it all} gas (assumed to trace the dark matter), not the warm-hot
baryons.  Such a clumping factor is dominated by the contribution from
galaxies and other collapsed, virialized structures.  This is not the
appropriate clumping factor with which to calculate the soft X-ray
background, since as we have shown in the previous section, WHIM gas is
typically at much lower overdensities than virialized structures.  In
\cite{pen99}, the clumping factor is strongly dependent on simulation
resolution, because it is dominated by the highest overdensity
objects.  Amongst our simulations, a wide range of resolutions give
barely an order of magnitude difference in clumping factor, since the
WHIM gas is predominantly diffuse.

\cite{pen99} determines that $\sim 1$~keV of non-gravitational heating
would reduce his clumping factor to $C\la 60$, which he argues is
necessary to satisfy XRB constraints.  However, this was derived
assuming that all gas contributes to the SXRB (see his eq.~2), whereas
in our simulations only $\sim 30\%$ of the baryons contribute to the SXRB
(probably an even smaller fraction since the narrow range of SXRB energies
arises from a narrower range of gas temperatures than $10^5<T<10^7$~K).
From equation~\ref{eqn: sxrb}, we see that a reduction in $\Omega_g$,
the mean density of emitting gas, results in an increase in the maximum allowed
clumping factor by $(\Omega_b/\Omega_{\rm WHIM})^2\approx 10$.  Thus his
analytically-derived constraint translates to $C_{\rm WHIM}\la 600$.
Our clumping factors ($\la 400$) are consistent with his analysis,
despite the fact that WHIM gas in our simulations undergoes almost no
non-gravitational heating.

\cite{wu99} present a more detailed model from which we can crudely
estimate our SXRB by comparison.  They do not specifically use clumping
factors to predict the SXRB in their model, but they perform a similar
calculation to that of \cite{pen99}, based on a \cite{pre74} analysis
with an assumed halo profile.  The \cite{pre74} formalism makes two
assumptions about how gas is distributed and heated in the universe:
(1) All soft X-ray emitting gas is bound in virialized halos (namely,
groups of galaxies); (2) Gas can only be heated by accretion onto a
virialized halo, with the gas temperature set by the virial temperature of the
halo.  Unlike \cite{pen99}, they allow some gas to cool and
therefore not emit in X-rays, but this is a small correction.  In their
model, the gas clumping factor is the clumping factor of {\it
virialized} objects, since they are assuming that emitting gas solely
resides in such objects.

Our simulations suggest that both \cite{pre74} assumptions are
strongly violated in the case of WHIM gas.  First, only a small
fraction of WHIM gas is bound in virialized halos; most is
distributed much more diffusely.  Second, in our simulations,
intergalactic gas can be heated significantly by purely gravitational
processes prior to being accreted onto a virialized object.

\PSbox{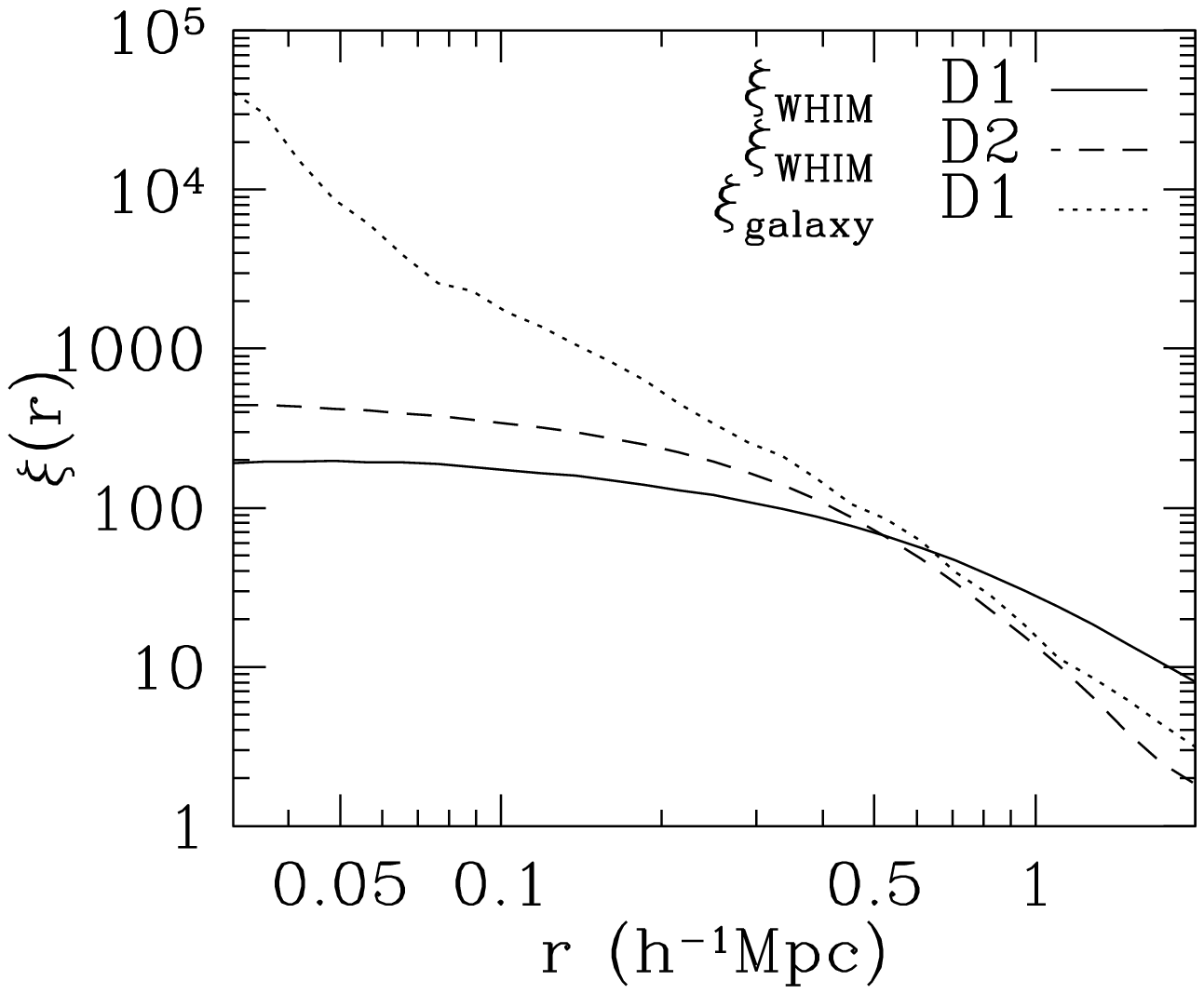 angle=0 voffset=-160 hoffset=-10 
vscale=40 hscale=40}{3.0in}{1.9in}
{\\\small Figure~7:  Two-point correlation functions for WHIM
gas from simulation D1 (solid line) and D2 (dashed line), and
for galaxies from simulation D1 (dotted line).
\vskip0.1in
}

The two-point correlation function $\xi_{\rm WHIM}$ illustrates quite
clearly that WHIM gas is not bound in virialized halos, and is instead
associated with large-scale structure.  Figure~7 shows $\xi_{\rm WHIM}$
for simulation D1 (solid line), D2 (dashed), as well as $\xi_{\rm
galaxy}$ (dotted line) calculated from D1.  $\xi_{\rm galaxy}$ is
computed as the correlation function of all star particles in this
simulation, where stars are formed slowly out of cold, dense gas and
are seen to trace the galaxy population (see \cite{kat96}).  At scales
less than a few hundred kpc, $\xi_{\rm WHIM}$ flattens, indicating that
WHIM gas does not cluster on scales smaller than those typical of
large-scale structures.  Conversely, the galaxy correlation
function continues as approximately a power law in to the resolution
limit of our simulation.  The clumping factor of WHIM gas, $\xi_{\rm
WHIM}(0)$, is well-defined since we can meaningfully extrapolate from
our resolution limit to $r=0$.  This is not true for the clumping
factor of galaxies, for which we may only set a lower limit (consistent
with Pen 1999), $C_{\rm galaxy}\ga 4\times 10^4$.

We can crudely estimate the SXRB flux we would predict in comparison to
that predicted by \cite{wu99} by multiplying their predicted $j_{\rm
SXRB}$ by $C_{\rm WHIM}/C_{\rm galaxy}$.  This corrects for the fact
that they use virialized objects (which we take as galaxies) instead
of the true diffuse WHIM responsible for SXRB emission.  Thus we would
roughly predict a SXRB flux that is two orders of magnitude lower than
that predicted by \cite{wu99}.  Examining their Figure~2 or 3 shows that
such a reduction in $j_{\rm SXRB}$ makes the SXRB contribution from
warm-hot gas consistent with observational limits.

In summary, the discrepancies in the predicted soft XRB versus previous
studies arise from their use of a \cite{pre74} analysis, or an
equivalent method that implicitly places emitting gas into virialized
objects, in order to study an intergalactic component of baryons that
is inherently much more diffuse.  These methods do not allow for
gravitational shock heating in unbound objects such as filaments, thus
they are forced to postulate significant non-gravitational heating to
make the emitting gas more diffuse.  In our scenario, most WHIM gas has
never fallen into virialized objects, though it is still heated
(almost) purely via gravitational processes.  Scaling previous analyses
by the typical overdensities found in our simulations suggests that the
properties of the predicted WHIM are consistent with SXRB constraints.
However, these scaling arguments are approximate at best, and not
enough to guarantee consistency.  More accurate calculations of X-ray
emission from these simulations are certainly warranted and may yield
interesting constraints on the WHIM component; these issues will be
addressed in detail in separate papers (Fardal et al., in
preparation; Phillips et al., in preparation).

\section{Summary}\label{sec: disc}

We study the warm-hot intergalactic medium (WHIM), defined as all the
gas in the universe with temperature $10^5<T<10^7$~K, in six cosmological
hydrodynamic simulations with widely varying spatial resolutions, volumes,
code algorithms, and input physics.  In each simulation, the WHIM contains
$\approx 30-40\%$ of all baryons in the present-day universe.  As a rule
of thumb, our simulations predict that the fractions of baryons in the
warm-hot phase, the diffuse phase, and gravitationally bound systems are
roughly comparable at the present epoch.

The WHIM is comprised primarily of gas at moderate overdensities, with
a median overdensity $\sim 10-30$,  and is predominantly an
intergalactic component.  It is shock-heated by accretion onto
non-equilibrium filamentary large-scale structures, with possibly a
small energy contribution from non-gravitational processes such as
supernova feedback.  Despite being heated primarily by gravitational
processes, WHIM gas in our simulations is consistent with constraints
from the soft X-ray background.  The clumping factors of WHIM gas in
our simulations range from $\sim 30-400$, which are far below that of
virialized objects ($\ga 10^4$) that some studies assume are the
sources of soft XRB photons.

Different simulations give somewhat different fractions of WHIM gas at
the present epoch, though the fraction is significant in all
simulations.  We argue that the differences are primarily attributable
to resolution effects, specifically manifested in the distribution of
supernova feedback energy in different simulations.  In high spatial
resolution simulations, the added heat remains in dense regions and
radiates away almost immediately, whereas in lower spatial resolution
simulations a significant fraction of heat ends up in diffuse
intergalactic gas where it cannot radiate away.  Still, the differences
in present-day WHIM fraction are not large between these two somewhat
extreme cases for feedback.  Radiative cooling plays a minor role in
the evolution of warm-hot gas because it is mostly at low densities,
and is too diffuse to self-gravitate and self-shield into mini-halos.
Other physical effects such as star formation, photoionization, and
cosmic variance do not produce significant differences in the amount or
properties of warm-hot gas in our simulations.

Since the exact predicted fraction of WHIM gas is as yet sensitive to
simulation details, it would be greatly beneficial to place
observational constraints on this component.  As mentioned before, this
is a challenging task.  Still, there are tantalizing hints of
detections of WHIM gas.  \cite{wan93} found an excess emission
component in ROSAT data around $T\sim 2\times 10^6$~K, which may be
arising from this diffuse component.  \cite{sol96} detected an
auto-correlation signature between the soft X-ray background and
galaxies, as would be expected if warm-hot gas was distributed in
large-scale structures.  Even more exciting is a possible direct
detection of diffuse soft X-ray emission associated with a filament of
galaxies, by \cite{sch00}.  Somewhat stronger evidence comes from a
census of \ion{O}{6} absorbers at redshifts $0.14\la z\la 0.27$
(\cite{tri00}; \cite{tri00b}), which together with conservative
ionization corrections and metallicities implies $\Omega_{\rm WHIM}\ga
0.003^{+0.004}_{-0.002} h_{75}^{-1}$, or $\Omega_{\rm WHIM}/\Omega_b
\ga 10$\% at the present epoch.  However, some of these \ion{O}{6}
absorbers may be photoionized, arising in cooler, low-density
intergalactic gas at $T\sim 10^4$~K.

WHIM gas emission may be easiest to detect around high density regions
such as clusters because that is where the density and temperature
are highest within the WHIM range (cf. Figure~6), but this is
{\it not} where the majority of WHIM gas is located.  For instance,
high-resolution Chandra spectra of regions between clusters that are free
of bright sources may provide a detectable signal (Phillips et al., in
preparation), and observations from XMM could possibly image filaments
of warm-hot gas directly (\cite{pie00}).  Our simulations, however,
predict the majority of WHIM gas is far away from galaxies and clusters,
residing in the diffuse IGM.  A promising avenue to detect this more
typical WHIM gas is via absorption, as continuing observations with STIS
aboard Hubble will detect many more \ion{O}{6} absorbers.  Future X-ray
satellites may be able to detect higher ionization absorbers such as
\ion{O}{7} and \ion{O}{8} that may also trace WHIM gas (\cite{hel98}).
The detection of this component is a key observational challenge, as the
warm-hot intergalactic medium is rapidly become an integral part of our
understanding of the evolution of baryons in the universe.

\bigskip
\acknowledgments

We thank Jeff Gardner, Ed Jenkins, Richard Mushotsky, Arielle Phillips, 
Jim Peebles and Todd Tripp for helpful discussions.
RD is supported by NASA ATP grant NAG5-7066.  
RC and JPO are supported by NSF grants AST-9803137 and ASC-9740300.
Support for GLB was provided by NASA through Hubble Fellowship grant
HF-01104.01-98A from the Space Telescope Science Institute, which is
operated by the Association of Universities for Research in Astronomy,
Inc., under NASA contract NAS5-26555.  
This work was supported by NASA Astrophysical Theory Grants NAG5-3922,
NAG5-3820, and NAG5-3111, by NASA Long-Term Space Astrophysics Grant
NAG5-3525, and by the NSF under grants ASC93-18185, ACI96-19019, and
AST-9802568.  
Some of the simulations were performed at the San Diego Supercomputer Center. 
We also thank NCSA for use of their computing facilities.

\end{document}